\documentclass[twocolumn,prb,showpacs,preprintnumbers,amsmath,amssymb]{revtex4}

\usepackage{subfigure}
\usepackage{graphicx}
\usepackage{dcolumn}
\usepackage{bm}
\usepackage{epsfig}

\begin{document}

\preprint{draft version 09/30/07}

\title{Profile of the U \textit{5f} magnetization in U/Fe multilayers}

\author{S. D. Brown}
\altaffiliation[Also at ]{Department of Physics, University of Liverpool, Liverpool L69 7ZE, United Kingdom}
\author{L. Bouchenoire}
\altaffiliation[Also at ]{Department of Physics, University of Liverpool, Liverpool L69 7ZE, United Kingdom}
\author{P. Thompson}
\altaffiliation[Also at ]{Department of Physics, University of Liverpool, Liverpool L69 7ZE, United Kingdom}
\affiliation{XMaS, UK-CRG, European Synchrotron Radiation Facility,
BP220, F-38043 Grenoble Cedex, France}

\author{R. Springell}
\altaffiliation[Also at ]{Department of Physics and Astronomy, University College London, London WC1E 6BT, United Kingdom}%
\author{A. Mirone}
\author{W. G. Stirling}
\altaffiliation[Also at ]{Department of Physics, University of Liverpool, Liverpool L69 7ZE, United Kingdom}
\affiliation{European Synchrotron Radiation Facility, BP220, F-38043
Grenoble Cedex, France}

\author{A. Beesley}
\author{M. F. Thomas}
\affiliation{Department of Physics, University of Liverpool,
Liverpool L69 7ZE, United Kingdom}

\author{R. C. C. Ward and M. R. Wells}
 \affiliation{Clarendon Laboratory, University of Oxford, Oxford, Oxon OX1 3PU, United Kingdom}%

\author{S. Langridge}
 \affiliation{Rutherford Appleton Laboratory, Chilton, Didcot, Oxon OX11 0QX, United Kingdom}%

\author{S. Zochowski}
 \affiliation{Department of Physics and Astronomy, University College London, London WC1E 6BT, United Kingdom}%

\author{G. H. Lander}
 \affiliation{European Commission, JRC, Institute for Transuranium Elements, Postfach 2340, 76125 Karlsruhe, Germany}%

\date{\today}

\begin{abstract}
Recent calculations, concerning the magnetism of uranium in the U/Fe
multilayer system have described the spatial dependence of the
\textit{5f} polarization that might be expected. We have used the
x-ray resonant magnetic reflectivity technique to obtain the profile
of the induced uranium magnetic moment for selected U/Fe multilayer
samples. This study extends the use of x-ray magnetic scattering for
induced moment systems to the \textit{5f} actinide metals. The
spatial dependence of the U magnetization shows that the predominant
fraction of the polarization is present at the interfacial
boundaries, decaying rapidly towards the center of the uranium
layer, in good agreement with predictions.
\end{abstract}

\maketitle

\section{\label{sec:level1}Introduction}

Magnetic multilayers exhibit a broad range of interesting phenomena,
which have both technological and scientific importance
\cite{Bland}. These properties are primarily driven by the
electronic interactions at the multilayer interfaces. Our interest
lies in the fundamental nature of interactions between the U
\textit{5f} and Fe \textit{3d} electrons, particularly concerning
the magnetism of uranium.

Theoretical predictions have been made regarding the polarization of
uranium in U/Fe multilayer systems, based on both scalar and fully
relativistic calculations \cite{Laref}. The approach adopted an
exchange correlation potential treated in the generalized gradient
approximation (GGA), which reproduced earlier theoretical
predictions for the magnetism of the surface of alpha uranium
\cite{Stojic}. A model U(001)/Fe(110) supercell structure was
proposed with lattice constants taken as the average between those
for uranium and iron. The calculations revealed the importance of
U-Fe electronic hybridization and predicted a spin moment on the
uranium site of $0.92\mu_{B}$, significantly larger and aligned
opposite to that of the orbital moment, $0.16\mu_{B}$. The total
uranium moment was predicted to align antiparallel to the Fe moments
and to decrease rapidly within just two atomic layers. We have
employed the x-ray resonant magnetic scattering technique in
reflection geometry to probe directly the spatial dependence of the
U polarization.

Since the first experimental evidence for x-ray magnetic scattering
\cite{deBergevin}, advances in x-ray sources and development of new
materials have lead to a surge of scientific activity in this field.
This has been aided by the discovery of large resonant enhancements
of the magnetic scattering at the $L_{2,3}$ edges of the rare earth
metals \cite{Gibbs} and the $M_{4,5}$ edges of the actinides
\cite{Isaacs, Tang}. In multi-component systems, such as magnetic
multilayers, resonant enhancements in the scattering factor can
dramatically improve the chemical contrast between elements.
Moreover, it is possible to detect strong magnetic dichroism at
these resonant absorption edges with the employment of polarized x
rays. X-ray resonant magnetic reflectivity (XRMR) combines the
benefits of magnetic dichroism with structural information from the
charge scattering so that it is possible to determine the spatial
profile of the magnetization within the layers. It is a technique
ideally suited to the investigation of the magnetism of uranium in
U/Fe multilayers. The use of XRMR to investigate buried interfaces
in magnetic nanostructures is well-documented, certainly for the
case of soft x rays \cite{Grabis}, but has been less exploited in
the hard x-ray regime. At these energies, it is most often the L
edges of the rare-earth elements which are of interest \cite{Seve},
but some progress has been made in understanding the induced moment
in \textit{5d} transition metal systems also \cite{Jaouen}.

Preliminary measurements at the U $M_{4}$ edge have reported an
induced magnetic moment on the U site \cite{Brown} and later, the
separation of spin and orbital components of this moment, using
x-ray magnetic circular dichroism (XMCD) \cite{Wilhelm}. The total
magnetic moments in this case are small ($\mathrm{\sim0.1
\mu_{B}}$). However, the element selectivity and brightness of x
rays at a synchrotron source, coupled with the large resonances of
the U \textit{M} edges, allow the moments to be easily detected. The
XMCD technique provides only an average of the U magnetization for
the whole multilayer and is not sensitive to the distribution of the
\textit{5f} polarization within the layer. The spatial dependence of
the induced U moment, determined by XRMR, provides a unique insight
into the extent of the U \textit{5f}, Fe \textit{3d} interaction.

\section{X-ray resonant magnetic reflectivity}

X-ray reflectivity measurements have become standard practise for
determining the structure of multilayers. Most commonly, a single
wavelength measurement is used and a calculation of the reflected
intensity, based on Parratt's recursive method \cite{Parratt} is
employed to model the layer thickness and roughness parameters.
However, in order to determine a profile of the magnetization of a
layer, particularly one whose polarization is strongly
thickness-dependent, it is important to determine more precisely the
interfacial structure. This can be achieved by varying the
electronic contrast of the respective elements; measuring the
scattered intensity as a function of energy through an absorption
edge of one of the constituent materials, in this case the $M_{4}$
edge of uranium.

\begin{figure}[htbp]
\centering
\includegraphics[width=0.4\textwidth,bb=180 340 430 540,clip]{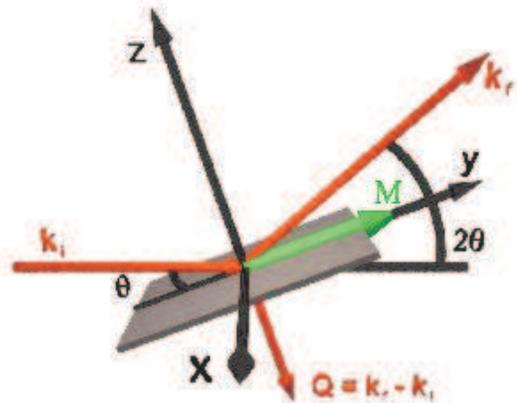}\caption
{\label{figure1}Schematic diagram of the longitudinal geometry used
in the XRMR measurements. $\mathbf{k_{i}}$ and $\mathbf{k_{f}}$ are
the wavevectors of the incoming and outgoing circularly polarized x
rays, respectively. $\mathbf{M}$ is the magnetization of the
sample.}
\end{figure}

Once the multilayer structure has been determined it is possible to
measure the $\mathbf{Q}$-dependent magnetic scattering, see Fig.
\ref{figure1}. By employing circularly polarized synchrotron
radiation and applying a magnetic field at the sample position, the
magnetic signal is detected as the difference in intensity of the
elastic scattering when either the helicity of the incoming x rays
or the magnetic field direction is reversed.

The atomic scattering factor, $F(E)$, can be written in terms of a
combination of charge, $F_{c}(E)$, and magnetic, $F_{m}(E)$,
structure factors \cite{Hannon}.

\begin{equation}\label{F}
F(E)=(\mathbf{\hat{\varepsilon}_{f}}\cdot\mathbf{\hat{\varepsilon}_{i}})F_{c}(E)-i(\mathbf{\hat{\varepsilon}_{f}}\times\mathbf{\hat{\varepsilon}_{i}})F_{m}(E)
\end{equation}

\noindent where $\mathbf{\hat{\varepsilon}_{i}}$ and
$\mathbf{\hat{\varepsilon}_{f}}$ are the unit polarization vectors
of the incident and scattered x rays, respectively. The charge
structure factor can be written as a summation over all the atoms in
the multilayer,

\begin{equation}
F_{c}=\sum(f_{0}+f'_{c}(E)+if''_{c}(E))e^{i\mathbf{Q}\cdot
\mathbf{r}}
\end{equation}

\noindent where $f_{0}$ is the tabulated atomic form factor
\cite{Henke}, and $f'_{c}(E)$ and $f''_{c}(E)$ are the real and
imaginary parts of the complex resonant anomalous scattering factor,
respectively. The resonant magnetic structure factor can be written
in a similar way, as a summation over the resonating magnetic atoms,

\begin{equation}
F_{m}=\sum
\mathbf{\hat{z}}(f'_{m}(E)+if''_{m}(E))e^{i\mathbf{Q}\cdot
\mathbf{r}}
\end{equation}

Here, $f'_{m}(E)$ and $f''_{m}(E)$ are the real and imaginary parts
of the resonant magnetic scattering factor, respectively, and
$\mathbf{\hat{z}}$ is the unit vector along the quantization axis
parallel to the local magnetic moment. The U $M_{4}$ absorption edge
represents the excitation of electrons from the 3$d_{3/2}$ to the
5$f_{5/2}$ states, where electric dipole transitions provide the
strongest contributions to the magnetic scattering
\cite{vanderLaan}. For dipole transitions, the resonant magnetic
scattering factors can be represented as,

\begin{equation}
f'_{m}(E)+if''_{m}(E)=\left(\frac{3}{4kr_{e}}\right)[F_{11}(E)-F_{1-1}(E)]
\end{equation}

\noindent where $F_{LM}$ is determined by atomic properties and is
related to the strength of the resonance, $r_{e}$ is the classical
electron radius and \textit{k} is the wavevector.

The intensities observed in elastic scattering are related to the
square of the atomic scattering factor, which on inspection of
equation (\ref{F}) yields cross terms that represent the resonant
magnetic-charge interference scattering.

\pagebreak

\begin{equation}\label{interference}
\sum|F^{+}|^{2}-\sum|F^{-}|^{2}=-2(\mathbf{\hat{k}}+\mathbf{\hat{k}'}\cos2\theta)\cdot(F'_{c}F'_{m}+F''_{c}F''_{m})
\end{equation}

\begin{equation}
=I^{+}-I^{-}
\end{equation}

The magnetic-charge interference can be accessed either by flipping
the helicity of the incoming photons or by flipping the magnetic
field. Conventionally + represents right circularly polarized (RCP)
x rays and $-$ represents left (LCP). In our case, we held the
polarization constant and flipped the magnetic field. Equation
(\ref{interference}) indicates that the magnetic-charge interference
scattering is only sensitive to the component of the magnetization
within the scattering plane, hence the magnetic field was applied
along the direction defined by the sample plane and the scattering
plane, see figure \ref{figure1}.

The scattered intensity was modeled by adapting Parratt's recursion
formula \cite{Parratt} for nonmagnetic specular reflectivity from a
multilayer. The complex amplitudes of the electric fields of the
transmitted and reflected x rays of both magnetic field states were
included in the calculations. In the frame of reference of the
incoming circularly polarized x-ray beam, the x rays experience
different refractive indices for each of the magnetic field
directions, $n^{\pm}=1-\delta^{\pm}+i\beta^{\pm}$.

\begin{equation}
\delta^{\pm}=\left(\frac{2\pi
n_{0}r_{e}}{k^{2}}\right)(f_{0}+f'_{c}(E)\mp
f'_{m}(E)\cos\theta\cos\phi)
\end{equation}

\begin{equation}
\beta^{\pm}=\left(\frac{2\pi n_{0}r_{e}}{k^{2}}\right)(f''_{c}(E)\mp
f''_{m}(E)\cos\theta\cos\phi)
\end{equation}

\noindent where $n_{0}$ is the number of atoms per unit volume. The
imaginary parts of the charge and magnetic scattering factors are
then modeled as a function of energy and a Kramers-Kronig
transformation is used to relate the real and imaginary parts of the
respective scattering factors.

Commonly, the optical parameters are well known, such as for the
transition metal L edges, and only a token number of energies need
be sampled to well describe the energy dependence of the scattered
intensity. Another approach is to measure the fluorescence and
calculate the XMCD, where the XMCD absorption coefficient is related
to the imaginary part of the magnetic scattering factor,

\begin{equation}
\mu_{m}(E)=-\left(\frac{8\pi
n_{0}r_{e}}{k}\right)(\mathbf{\hat{k}}\cdot\mathbf{\hat{z}})f''_{m}(E)
\end{equation}

However, significant self absorption effects can be present in
fluorescence measurements \cite{Pfalzer}. The corrections for these
effects then presuppose a knowledge of the structure. In our case
the optical constants are not well known and the resonance of the U
$M_{4}$ edge is large. Also, previous measurements on the U/Fe
system \cite{Beesley1, Beesley2, Springell} have indicated that the
structure of the U/Fe multilayer interfaces cannot be modeled
simply. In this investigation we have used a double Lorentzian
squared line-shape to model the imaginary part of the scattering
factor and an arctan function to model the non-resonant
photoelectric absorption. A total of 17 energies were used to
precisely track the scattered intensity as a function of energy.

The layer was divided into slices along the z-direction,
approximately one atomic plane in thickness ($\mathrm{\sim2.5\AA}$).
The interfacial structure was then modeled by varying the relative
densities of the uranium and iron to give a profile of both the U
and Fe densities through the multilayer. The calculation of the
charge scattering was fitted to the experimental data simultaneously
for two different multilayers at all energies. Several parameters
were held constant from sample to sample. An important
simplification included using identical interfacial regions for each
bilayer for each sample.

In order to determine a profile of the induced U magnetization a
coefficient was applied to the magnetic scattering factors for each
slice of the bilayer that contained some uranium density. The
coefficients are proportional to the magnetic moment per uranium
atom, hence, by fitting these values to the magnetic-charge
interference scattering it is possible to model the spatial
dependence of the induced U \textit{5f} polarization along the
growth direction.

\section{Experimental Details}

The samples were prepared by dc magnetron sputtering in a UHV
loadlocked growth chamber, operating at a base pressure of
$5\times10^{-10}\mathrm{mbar}$. The multilayers were grown on
$\mathrm{50\AA}$ thick niobium buffer layers deposited onto
single-crystal sapphire plates. The multilayers were sputtered at a
growth rate of $\mathrm{\sim1\AA/s}$ in an argon atmosphere of
$\mathrm{5\times10^{-3}mbar}$. The samples were protected from
oxidation by a $\mathrm{50\AA}$ Nb capping layer. The structural and
bulk magnetic properties have been reported previously
\cite{Springell, Springell2}.

The XRMR measurements were carried out at the XMaS beamline (BM28)
at the ESRF in Grenoble. This beamline is situated on a bending
magnet section of the synchrotron, where the optics and experimental
hutch set-up have been designed for the study of x-ray magnetic
scattering and the photon flux has been optimized at energies in the
vicinity of the U \textit{M} edges. A complete description of the
beamline optics and experimental capabilities has been reported
\cite{Brown2}. The sample views the x-ray beam on orbit, so that the
incident flux is linearly polarized. A 90\% rate of circular
polarization was achieved, by employing a diamond (111),
quarter-wave, phase-plate \cite{Bouchenoire}. In order to preserve
as much flux as possible, necessary at the relatively low energies
of the uranium \textit{M} edges, the flight paths of the incident
and scattered x rays were under high vacuum.

The samples were mounted on copper stubs and attached to a magnet
assembly, consisting of water-cooled pole pieces, generating an
applied magnetic field of 0.1T. This field was large enough to
saturate the iron moments, but small enough to be flipped rapidly.
The pole pieces were arranged so that they could provide a field
aligned parallel to the scattering plane. The magnet was fixed on a
precision sample mount on an 11 circle Huber diffractometer and a
Bicron detector was mounted on the $\mathrm{2\theta}$ arm. All
experiments were performed at room temperature, as the previous
studies showed a significant dichroic signal up to 300 K
\cite{Wilhelm}.

The measurements were carried out on two U/Fe samples, SN71,
$\mathrm{[U_{9}/Fe_{34}]_{30}}$ and SN76,
$\mathrm{[U_{27}/Fe_{57}]_{20}}$, whose nominal layer thicknesses
were determined by x-ray reflectivity \cite{Springell}. XRMR
measurements were made across the U $M_{4}$ edge (3728 eV) at room
temperature, for circularly polarized x rays in an applied field of
0.1T. The field was flipped in the following sequence: $+ - - +$, at
each point. Data were collected for 17 energies spanning 20 eV below
the $M_{4}$ edge to 20 eV above, providing a mesh of the x-ray
reflectivity and magnetic-charge interference scattering (defined as
$\mathrm{I^{+}-I^{-}}$) as a function of $\mathbf{Q}$ and energy.

\section{Results}

\begin{figure}[htbp]
\centering
\includegraphics[width=0.45\textwidth,bb=10 10 260 285,clip]{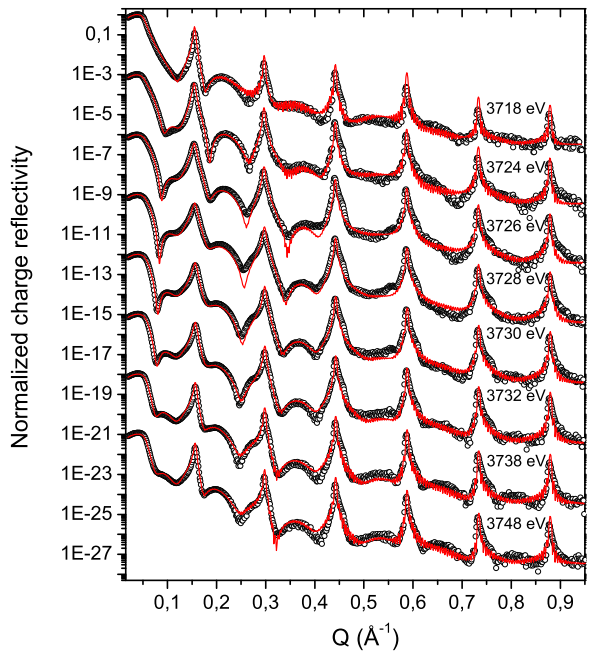}\caption
{\label{figure2}Energy variation of the charge reflectivity in the
vicinity of the U $M_{4}$ edge. The experimental data are shown as
the open black circles and the calculated reflectivity is
represented by the solid red line. The spectra have been scaled by
factors of $\mathrm{10^{-3}}$ for clarity.}
\end{figure}

\begin{figure}[htbp]
\centering
\includegraphics[width=0.4\textwidth,bb=5 5 230 290,clip]{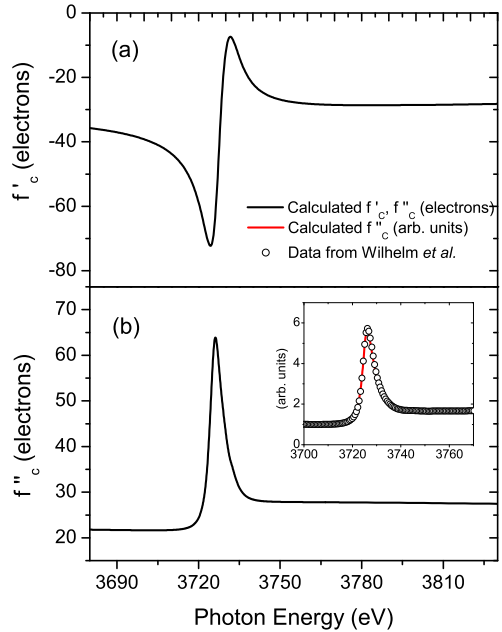}\caption
{\label{figure3}The real (a) and imaginary (b) parts of the resonant
scattering factor are shown in electron units. The imaginary part is
determined from the calculations of the charge reflectivity and the
real part is its Kramers-Kronig transform. The insert of (b) is a
comparison of the modeled imaginary scattering factor and the
fluorescence, reported previously \cite{Wilhelm}.}
\end{figure}

\begin{figure}[htbp]
\centering
\includegraphics[width=0.37\textwidth,bb=10 10 210 285,clip]{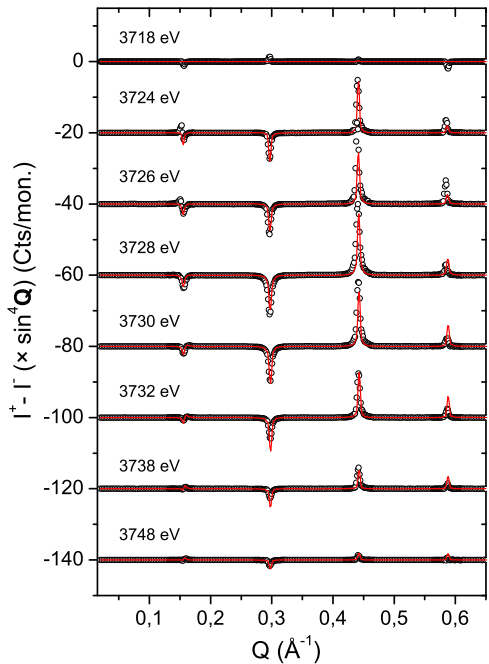}\caption
{\label{figure4}The magnetic-charge interference scattering as a
function of energy across the U $M_{4}$ edge. The data are shown as
the open black circles and the fitted calculation is represented by
the solid red line, both are scaled by a
$\mathrm{\sin^{4}\mathbf{Q}}$ factor. The results presented for each
energy have been offset by 20 cts/mon for clarity.}
\end{figure}

\begin{figure}[htbp]
\centering
\includegraphics[width=0.37\textwidth,bb=5 5 230 290,clip]{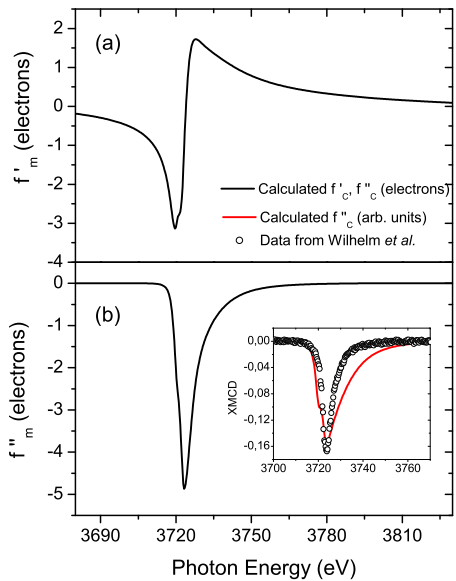}\caption
{\label{figure5}The real (a) and imaginary (b) parts of the resonant
magnetic scattering factor. The imaginary part is determined from
the calculations of the charge-magnetic interference scattering and
the real part is its Kramers-Kronig transform. The insert of (b) is
a comparison of the XMCD signals determined by this model and
measurements made at room temperature in a field of 1T by Wilhelm
\textit{et al} \cite{Wilhelm}.}
\end{figure}

Fig. \ref{figure2} shows the experimental x-ray resonant
reflectivity data and calculated intensities for sample SN71. The
model reproduces well the fine detail contained between the Bragg
peaks and the changes in the reflectivity with energy. A particular
feature is the broadening of the Bragg peaks at the resonant energy.
Fig. \ref{figure3} shows the real (a) and imaginary (b) parts of the
resonant scattering factor. The imaginary part was determined from
the model described in section II and the real part was taken as its
Kramers-Kronig transform, both shown in electron units. The insert
of Fig. \ref{figure3} (b) shows a comparison of the self-absorption
corrected fluorescence data \cite{Wilhelm} with the imaginary part
of the resonant scattering factor.

The magnetic-charge interference scattering,
$\mathrm{(I^{+}-I^{-})}$, was calculated separately for each sample.
A real and imaginary magnetic scattering factor was modeled for each
sample. Fig. \ref{figure4} shows the magnetic difference across the
first four Bragg peaks in the vicinity of the U $M_{4}$ edge for
sample SN71, compared to the fitted calculation. Higher order Bragg
peaks did not yield a measurable magnetic effect. The experimental
data and the calculated intensities have been scaled by the
theoretical \textbf{Q}-dependence of the reflected intensity.

The imaginary part of the magnetic scattering factor used to model
the magnetic-charge interference scattering for sample SN71 is shown
in Fig. \ref{figure5} (b) and the real part (a) is its
Kramers-Kronig transform. The insert of Fig. \ref{figure5} (b) shows
a comparison between the XMCD signal determined from the imaginary
part of the magnetic scattering factor (normalized to the imaginary
part of the charge scattering factor) and XMCD data (normalized to
the fluorescence) measured on the same sample at room temperature
and in an applied field of 1T \cite{Wilhelm}.

\section{Discussion}

The agreement between calculated and experimental data is good for
both structural and magnetic data. The scattering factors of Fig.
\ref{figure3}, determined from the fitted calculations of the charge
scattering, are similar in shape and magnitude to those measured
across the U $M_{4}$ edge in other uranium systems \cite{Watson}.
The good agreement between the fluorescence measurements, corrected
for self absorption effects \cite{Wilhelm}, and the imaginary part
of the resonant scattering factor, shown in Fig. \ref{figure3} (b),
may be attributed to the large number of energies sampled. This is
also supported by the comparison of the XMCD signals shown in the
insert of Fig. \ref{figure5} (b).

\begin{figure*}[htbp]
\centering
\includegraphics[width=0.9\textwidth,bb=15 15 295 215,clip]{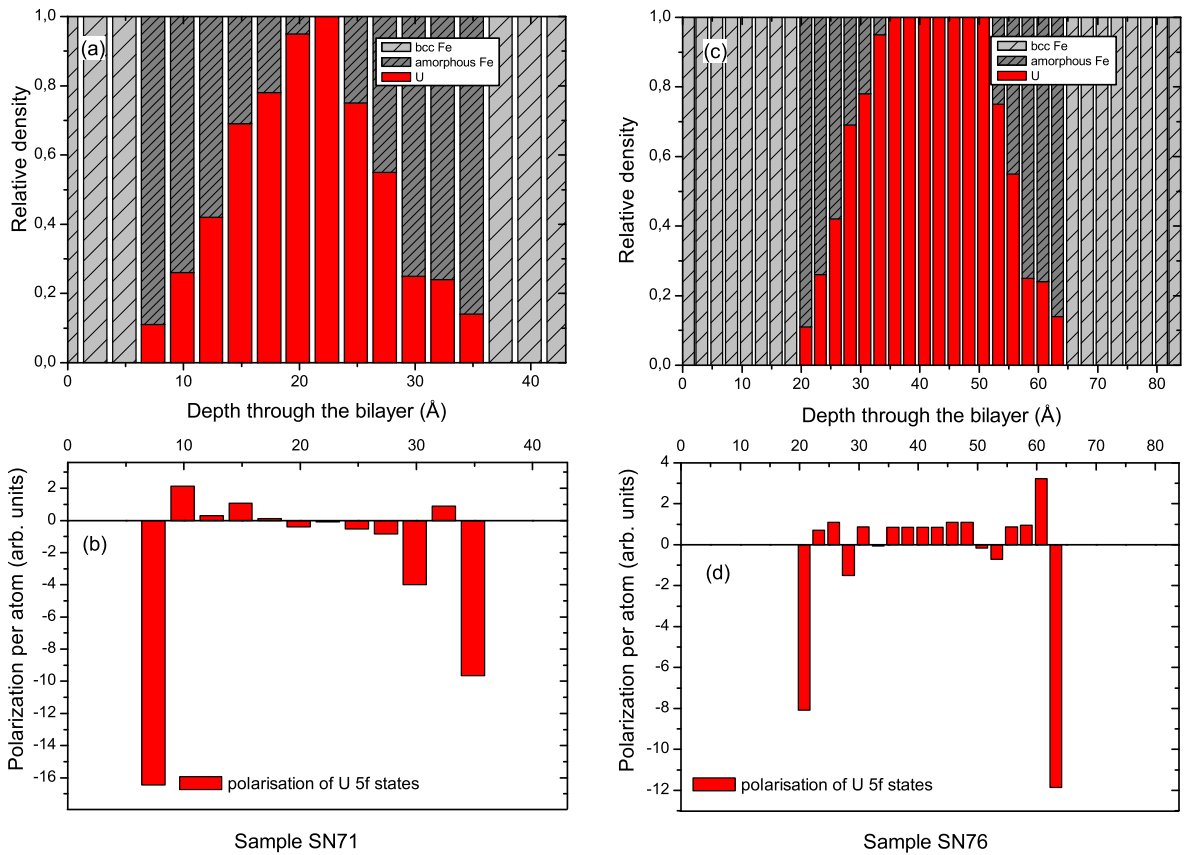}\caption
{\label{figure6}Profiles of the relative uranium and iron densities
as a function of bilayer depth are shown in panels (a) and (c) of
Fig. \ref{figure6} for samples SN71 and SN76, respectively. Panels
(b) and (d) present the profiles of the uranium polarization. We
note that the dominant negative sign of the polarization is an
assumption, based on XMCD measurements of these samples
\cite{Wilhelm}.}
\end{figure*}

The relative densities of the uranium and iron within a bilayer are
shown in panels (a) and (c) of Fig. \ref{figure6} for samples SN71
and SN76, respectively. The density profiles show extended regions
on either side of the central iron and uranium layers, which consist
of a mixture of iron and uranium atoms. This result is consistent
with the models proposed in earlier studies, using the M\"{o}ssbauer
technique \cite{Beesley2} and suggests that the non-magnetic "dead"
layer \cite{Springell2} could be a result of the alloying of the Fe
atoms in the interfaces, labeled as amorphous Fe in Fig.
\ref{figure6}. These alloy-type regions account for effects of
interfacial roughness and interdiffusion that can dramatically alter
the profile of the magnetization \cite{Sipr}.

The magnetic-charge scattering was calculated by assigning a
magnetization to each slice of the bilayer containing uranium. The
calculations were fitted to the experimental data without
restrictions on the shape or symmetry of the profile. The resultant
profiles of the induced magnetization within the uranium component
of the multilayers are shown in panels (b) and (d) of Fig.
\ref{figure6} and are scaled to the relative densities of uranium in
each slice. It is clear that the polarization occurs mainly when the
uranium atoms are close to the central iron layers, which contain
the magnetic bcc component. Furthermore, the magnetization of the
uranium falls off very rapidly away from the central iron layers and
is in the same direction at each side of the interface.

Calculations of the polarization of uranium in U/Fe multilayers used
a model system, which consisted of a sharp interface region and a
lattice-matched superstructure \cite{Laref}. A moment of about
$\mathrm{1\mu_{B}}$ was predicted, but was found to be considerably
smaller ($\mathrm{\sim0.1\mu_{B}}$)in XMCD studies \cite{Wilhelm}.
This can be attributed to differences between the idealized model
calculation and the real multilayers. The spatial dependence of the
induced magnetic moment was predicted to fall away very quickly from
the maximum value, so that within two atomic planes
($\mathrm{\sim5\AA}$) it is almost zero. In this respect the
calculations and experimental data are in good agreement.

\section{Conclusions}

Energy dependent x-ray resonant reflectivity at the U $M_{4}$ edge
has been used to determine the detailed structure of U/Fe
multilayers. Interfacial regions are present, containing a
uranium-iron alloy. The fitted charge scattering factors shown in
Fig. \ref{figure3} are similar to those found in other uranium
systems \cite{Watson} and are in close agreement with self
absorption corrected fluorescence data \cite{Wilhelm}.

The magnetic-charge interference scattering has been used to
determine the profile of the uranium magnetization. Agreement
between calculation and experiment could only be achieved with the
introduction of extended interdiffused regions at the interfacial
boundaries. The U polarization is predominantly at the low uranium
concentration end of the interface, in close proximity to the bcc
iron and decays rapidly as a function of depth towards the centre of
the uranium layer. This is in qualitative agreement with theoretical
calculations \cite{Laref}, emphasizing the importance of the
\textit{3d-5f} hybridization for the induced magnetization of
uranium in U/Fe multilayers.

\section{Acknowledgements}

R.S. acknowledges the receipt of an EPSRC research studentship.

\bibliography{UFe_XRMR}
\bibliographystyle{apsrev}

\end{document}